\documentclass{ws-ijmpd}
\newcommand{\beq}{\begin{equation}}
\newcommand{\eeq}{\end{equation}}
\newcommand{\beqn}{\begin{eqnarray}}
\newcommand{\eeqn}{\end{eqnarray}}

\newcommand{\n}{\mathbf{n}}
\newcommand{\p}{\mathbf{p}}

\newcommand{\om}{\mbox{{\boldmath$\omega$}}}

\newcommand{\na}{\mbox{{\boldmath$\nabla$}}}
\newcommand{\si}{\mbox{{\boldmath$\sigma$}}}

\newcommand{\al}{\mbox{${\alpha}$}}
\newcommand{\ga}{\mbox{${\gamma}$}}
\newcommand{\ep}{\mbox{${\varepsilon}$}}

\newcommand{\pa}{\mbox{${\partial}$}}
\begin{document}

\markboth{I.B. Khriplovich} {BIG BOUNCE AND INFLATION FROM
GRAVITATIONAL FOUR-FERMION INTERACTION}

%
\catchline{}{}{}{}{}
%

\title{BIG BOUNCE AND INFLATION\\
\vspace{2mm}
FROM GRAVITATIONAL FOUR-FERMION INTERACTION}

\author{I.B. KHRIPLOVICH}

\address{Budker Institute of Nuclear Physics\\
11 Lavrentjev pr., 630090 Novosibirsk, Russia\\
khriplovich@inp.nsk.su}

\maketitle

\begin{history}
\received{Day Month Year}
\revised{Day Month Year}
\comby{Managing Editor}
\end{history}

\begin{abstract}
The four-fermion gravitational interaction is induced by torsion,
and gets dominating on the Planck scale. The regular, axial-axial
part of this interaction by itself does not stop the gravitational
compression. However, the anomalous, vector-vector interaction
results in a natural way both in big bounce and in inflation.
\end{abstract}


\vspace{2mm}

1. The observation that, in the presence of torsion, the
interaction of fermions with gravity results in the four-fermion
interaction of axial currents, goes back at least to Ref. 1.

We start our discussion of the four-fermion gravitational
interaction with the analysis of its most general form.

As has been demonstrated in Ref. 2, the common action for the
gravitational field can be generalized as follows:
\beq\label{sg}
\noindent S_g = -\frac{1}{16\pi G}\int d^4 x\,(-e)\,e^{\mu}_I
e^{\nu}_J\left( R^{IJ}_{\mu\nu}
-\frac{1}{\ga}\tilde{R}^{IJ}_{\mu\nu}\right);
\eeq
\noindent here and below $G$ is the Newton gravitational constant,
$I,J=0,1,2,3$ (and $M,N$ below) are internal Lorentz indices,
$\mu,\nu=0,1,2,3$ are space-time indices, $e_\mu^I$ is the tetrad
field, $e$ is its determinant, and $e^{\mu}_I$ is the object
inverse to $e_\mu^I$. The curvature tensor is
\[
R^{IJ}_{\mu\nu}=-\pa_{\mu} \om^{IJ}{}_\nu+\pa_{\nu}\om^{IJ}{}_\mu+
\om^{IK}{}_{\mu}\,\om_K {}^{J}{}_\nu
-\om^{IK}{}_\nu\,\om_K{}^J{}_\mu,
\]
here $\om^{IJ}_{\mu}$ is the connection. The first term in
(\ref{sg}) is in fact the common action of the gravitational field
written in tetrad components.

The second term in (\ref{sg}), that with the dual curvature tensor
\[
\tilde{R}^{IJ}_{\mu\nu}=
\,\frac{1}{2}\,\varepsilon^{IJ}_{KL}R^{KL}_{\mu\nu}\,,
\]
does not vanish in the presence of spinning particles generating
torsion.

As to the so-called Barbero-Immirzi parameter $\ga$, its numerical
value
\beq
\ga = 0.274
\eeq
was obtained for the first time in Ref. 3, as the solution of the
"secular" equation
\beq
\sum_{j=1/2}^{\infty} (2j+1) e^{- 2\pi\gamma \sqrt{j(j+1)}} = 1.
\eeq

Interaction of fermions with gravity results, in the presence of
torsion, in the four-fermion action which looks as follows:
\beq\label{ff}
S_{ff}=\frac{3}{2}\pi G\frac{\ga^2}{\ga^2 + 1}\int d^4x\sqrt{-g}
\left[\,\eta_{IJ}A^I A^J + \frac{\al}{\ga} \eta_{IJ}(V^I A^J + A^I
V^J) -\al^2\,\eta_{IJ}V^I V^J\right];
\eeq
here and below $g$ is the determinant of the metric tensor, $A^I$
and $V^I$ are the total axial and vector neutral currents,
respectively:
\beq\label{A}
A^I = \sum_a A_a^I = \sum_a
\bar{\psi}_a\,\ga^5\,\ga^I\,\psi_a\,;\hspace{10mm} V^I = \sum_a
V_a^I = \sum_a \bar{\psi}_a\,\ga^I\,\psi_a\,;
\eeq
the sums over $a$ in (\ref{A}) extend over all sorts of elementary
fermions with spin 1/2.

The $AA$ contribution to expression (\ref{ff}) corresponds (up to
a factor) to the action derived long ago in Ref. 1. Then, this
contribution was obtained in the limit $\ga \to \infty$ in Ref. 4
(when comparing the corresponding result from Ref. 4 with
(\ref{ff}), one should note that our convention $\eta_{IJ} = {\rm
diag}(1, -1, -1, -1)$ differs in sign from that used in Ref. 4).
The present form of the $AA$ interaction, given in (\ref{ff}), was
derived in Ref. 5.

As to $VA$ and $VV$ terms in (\ref{ff}), they were derived in Ref.
6 as follows. The common action for fermions in gravitational
field
\beq
S_f = \int d^4 x\,\sqrt{-g}\,\frac{1}{2}\,[
 \,\bar{\psi}\,\ga^I\,e^{\mu}_I\,i\na_\mu \psi
 - \,i\,
 \overline{\na_\mu\psi}\,\ga^{I}e_{I}^\mu\psi]
\eeq
can be generalized to:
\begin{eqnarray}\label{sf}
S_f &=& \int d^4 x \,\sqrt{-g}\, \frac{1}{2}\,[
 (1-i\al)\,\bar{\psi}\,\ga^I\,e^{\mu}_I\,i\na_\mu \psi
 - (1+i\al)\,i\,
 \overline{\na_\mu\psi}\,\ga^{I}e_{I}^\mu\psi];
\end{eqnarray}
here
 $\na_\mu = \pa_\mu -
 \frac{1}{4}\,\om^{IJ}_{\;\;\;\;\mu}\,\ga_I\ga_J\,
$;\hspace{2mm} $\om^{IJ}_{\;\;\;\;\mu}$ is the connection. The
real constant $\al$ introduced in (\ref{sf}) is of no consequence,
generating only a total derivative, if the theory is torsion free.
However, in the presence of torsion this constant gets operative.
In particular, as demonstrated in Ref. 6, it generates the $VA$
and $VV$ terms in the gravitational four-fermion interaction
(\ref{ff}). 

Simple dimensional arguments demonstrate that interaction
(\ref{ff}), being proportional to the Newton constant $G$ and to
the particle number density squared, gets essential and dominates
over the common interactions only at very high densities and
temperatures, i.e. on the Planck scale.

The list of papers where the gravitational four-fermion
interaction is discussed in connection with cosmology, is too
lengthy for this short note. Therefore, I refer here only to the
most recent Ref. 7, with a quite extensive list of references.
However, in all those papers the discussion is confined to the
analysis of the axial-axial interaction.

In particular, in my paper Ref. 8 it was claimed that $VA$ and
$VV$ terms in formula (\ref{ff}) are small as compared to the $AA$
one. The argument was as follows. Under these extreme conditions,
the number densities of both fermions and antifermions increase,
due to the pair creation, but the total vector current density
$V^I$ remains intact.

By itself, this is correct. However, the analogous line of
reasoning applies to the axial current density $A^I$. It is in
fact the difference of the left-handed and right-handed axial
currents: $A^I = A^I_L - A^I_R$. There is no reason to expect that
this difference changes with temperature and/\,or pressure.

Moreover, the fermionic number (as distinct from the electric
charge) is not a long-range charge. Therefore, even the
conservation of fermionic number could be in principle violated,
for instance, by the decay of a neutral particle (majoron) into
two neutrinos. (I am grateful to A.D. Dolgov for attracting my
attention to this possibility.)

So, we work below with both currents, $A$ and $V$.

\vspace{3mm}

2. Let us consider the energy-momentum tensor (EMT) $T_{\mu\nu}$
generated by action (\ref{ff}). Therein, the expression in square
brackets has no explicit dependence at all either on the metric
tensor, or on its derivatives. The metric tensor enters action
$S_{ff}$ via $\,\sqrt{-g}$ only, so that the corresponding EMT is
given by relation
\beq
\frac{1}{2}\;\sqrt{-g}\;T_{\mu\nu} = \frac{\delta}{\delta
g_{\mu\nu}} S_{ff}\,.
\eeq
Thus, with identity
\beq
\frac{1}{\sqrt{-g}}\frac{\partial\sqrt{-g}}{\partial g^{\mu\nu}} =
- \frac{1}{2}\, g_{\mu\nu}\,,
\eeq
we arrive at the following expression for the EMT:
\[
T_{\mu\nu} = - \frac{3\pi}{2} \frac{\ga^2}{\ga^2 + 1}\,
g_{\mu\nu}\left[\,\eta_{IJ}A^I A^J
\,+\,\frac{\al}{\ga}\;\eta_{IJ}\,(V^I A^J + A^I V^J) -\al^2\,
\eta_{IJ}V^I V^J\,\right],
\]
or, in the tetrad components,
\beq\label{emt}
\noindent T_{MN} = - \frac{3\pi}{2} \frac{\ga^2}{\ga^2 + 1}\,
\eta_{MN}\left[\,\eta_{IJ}A^I A^J\,+
\,\frac{\al}{\ga}\;\eta_{IJ}\,(V^I A^J + A^I V^J) -\al^2\,
\eta_{IJ}V^I V^J\,\right].
\eeq
We note first of all that this EMT in the locally inertial frame
corresponds to the equation of state
\beq\label{pep}
p = -\,\ep;
\eeq
here and below $\ep = T_{00}$ is the energy density, and $p =
T_{11} = T_{22} = T_{33}$ is the pressure.

Let us analyze the expressions for $\ep$ and $p$ in our case of
the interaction of two ultrarelativistic fermions (labeled $a$ and
$b$) in their locally inertial center-of-mass system.

The axial and vector currents of fermion $a$ are, respectively,
\[
A^I_a = \frac{1}{4E^2}\;\phi^\dagger_a\,\{E\, \si_a\,(\p^\prime +
\p),\;(E^2  - (\p^\prime \p))\,\si_a + \p^\prime(\si_a\p) +
\p\,(\si_a \p^\prime) - i\,[\p^\prime \times \p]\}\,\phi_a =
\]
\beq
= \frac{1}{4}\;\phi^\dagger_a\,\{\si_a\,(\n^\prime + \n),\;(1 -
(\n^\prime \n))\,\si_a
 + \n^\prime(\si_a\n) + \n\,(\si_a \n^\prime) - i \,[\n^\prime \times \n]\}\,\phi_a;
\eeq
\[
V^I_a = \frac{1}{4E^2}\;\phi^\dagger_a\,\{E^2 + (\p^\prime \p) +
i\,\si_a\,[\p^\prime \times \p],\; E\left(\p^\prime + \p -
i\,\si_a \times (\p^\prime - \p)\right)\}\,\phi_a =
\]
\beq
= \frac{1}{4}\;\phi^\dagger_a\,\{1+(\n^\prime \n)
+i\,\si_a\,[\n^\prime \times \n],\;\n^\prime + \n \,- i\,\si_a
\times (\n^\prime - \n)]\}\,\phi_a;
\eeq
here $E$ is the energy of fermion $a$, $\n$ and $\n^\prime$ are
the unit vectors of its initial and final momenta $\p$ and
$\p^\prime$, respectively; under the discussed extreme conditions
all fermion masses can be neglected. In the center-of-mass system,
the axial and vector currents of fermion $b$ are obtained from
these expressions by changing the signs: $\n \to -\n$, $\n^\prime
\to -\n^\prime$. Then, after averaging over the directions of $\n$
and $\n^\prime$, we arrive at the following semiclassical
expressions for the nonvanishing components of the energy-momentum
tensor, i.e. for the energy density $\ep$ and pressure $p$ (for
the correspondence between $\ep$, $p$ and EMT components see Ref.
9, \S\;35):
\[
\ep = T_{00} = -\,\frac{\pi}{48}\,\frac{\ga^2}{\ga^2 +
1}\;G\,\sum_{a,\;b}\rho_a \,\rho_b\,[(3 - 11\,<\si_a\si_b>) -
\al^2(60 - 28\,<\si_a\si_b>)]
\]
\beq\label{ep}
=\,-\,\frac{\pi}{48}\,\frac{\ga^2}{\ga^2 + 1}\;G\,\rho^2\,[(3 -
11\,\zeta) - \al^2(60 - 28\,\zeta)]\,;
\eeq
\[
p = T_{11}= T_{22}= T_{33} = \frac{\pi}{48}\,\frac{\ga^2}{\ga^2 +
1}\;G\,\sum_{a,\;b}\rho_a \,\rho_b\,[(3 - 11\,<\si_a\si_b>) 
\]
\[
 - \,\al^2(60 - 28\,<\si_a\si_b>)]
\]
\beq\label{p}
=\,\frac{\pi}{48}\,\frac{\ga^2}{\ga^2 + 1}\;G\,\rho^2\,[(3 -
11\,\zeta) - \al^2(60 - 28\,\zeta)]\,;
\eeq
here and below $\rho_a$ and $\rho_b$ are the number densities of
the corresponding sorts of fermions and antifermions, $\rho =
\sum_a \rho_a$ is the total density of fermions and antifermions,
the summation $\sum_{a,\,b}$ extends over all sorts of fermions
and antifermions; $\zeta = \,<\si_a\si_b>$ is the average value of
the product of corresponding $\si$-matrices, presumably universal
for any $a$ and $b$. Since the number of sorts of fermions and
antifermions is large, one can neglect here for numerical reasons
the contributions of exchange and annihilation diagrams, as well
as the fact that if $\si_a$ and $\si_b$ refer to the same
particle, $<\si_a\si_b> = 3$. The parameter $\zeta$\,, just by its
physical meaning, in principle can vary in the interval from 0
(which corresponds to the complete thermal incoherence or to the
antiferromagnetic ordering) to 1 (which corresponds to the
complete ferromagnetic ordering).

It is only natural that after the performed averaging over $\n$
and $\n^\prime$, the $P$-odd contributions of $VA$ to $\ep$ and
$p$ vanish.

\vspace{3mm}

3. Though for $\al \sim 1$ the $VV$ interaction dominates
numerically the results (\ref{ep}) and (\ref{p}), it is
instructive to start the analysis with the discussion of the case
$\al = 0$, at least, for the comparison with the previous
investigations. We note in particular that, according to
(\ref{ep}), the contribution of the gravitational spin-spin
interaction to energy density is positive, i.e. the discussed
interaction is repulsive for fermions with aligned spins. This our
conclusion agrees with that made long ago in Ref. 4.

To simplify the discussion, we confine from now on to the region
around the Planck scale, so that one can neglect effects due to
the common fermionic EMT, originating from the Dirac Lagrangian
and linear in the particle density $\rho$.

A reasonable dimensional estimate for the temperature $\tau$ of
the discussed medium is
\beq\label{e3}
\tau \sim \,m_{\rm {Pl}}\
\eeq
(here and below $m_{\rm {Pl}}$ is the Planck mass). This
temperature is roughly on the same order of magnitude as the
energy scale $\omega$ of the discussed interaction
\beq\label{om}
\omega \sim \,G\,\rho\,\sim \,m_{\rm {Pl}}\,.
\eeq
Numerically, however, $\tau$ and $\omega$ can differ essentially.
Both options, $\tau > \omega$ and $\tau < \omega$, are
conceivable.

If the temperature is sufficiently high, $\tau \gg \omega$, it
destroys the spin-spin correlations in formulas (\ref{ep}) and
(\ref{p}). In the opposite limit, when $\tau \ll \omega$, the
energy density (\ref{ep}) is minimized by the antiferromagnetic
spin ordering. Thus, in both these limiting cases the energy
density and pressure simplify to
\beq\label{ep0}
\ep =  -\,\frac{\pi}{16}\,\frac{\ga^2}{\ga^2 + 1}\;G\,\rho^2;
\hspace{5mm} p = \frac{\pi}{16}\,\frac{\ga^2}{\ga^2 + 1}\;G \,
\rho^2\,.
\eeq
\noindent The energy density $\ep$, being negative and
proportional to $\rho^2$, decreases with the growth of $\rho$. On
the other hand, the common positive pressure $p$ grows together
with $\rho$. Both these effects result in the compression of the
fermionic matter, and thus make the discussed system unstable.

A curious phenomenon could be possible if initially the
temperature is sufficiently small, $\tau < \omega$, so that
equations (\ref{ep0}) hold. Then the matter starts compressing,
its temperature increases, and the correlator $\zeta =
\,<\si_a\si_b>$ could arise. When (and if!) $\zeta$ exceeds its
critical value $\zeta_{cr} = 3/11$, the compression changes to
expansion. Thus, we would arrive in this case at the big bounce
situation.

However, I am not aware of any physical mechanism which could
result here in the transition from the initial antiferromagnetic
ordering to the ferromagnetic one with positive $\,\zeta =
\,<\si_a\si_b>$.

Here one should mention also quite popular idea according to which
the gravitational collapse can be stopped by a positive spin-spin
contribution to the energy. However, how such spin-spin
correlation could survive under the discussed extremal conditions?
The na\"{\i}ve classical arguments do not look appropriate in this
case.

\vspace{3mm}

4. Let us come back now to equations (\ref{ep}), (\ref{p}). In
this general case, with nonvanishing anomalous $VV$ interaction,
the big bounce takes place if the energy density (\ref{ep}) is
positive (and correspondingly, the pressure (\ref{p}) is
negative). In other words, the anomalous, VV interaction results
in big bounce under the condition
\beq\label{al}
\al^2 \geq \,\frac{3-11\zeta}{4\,(15-7\zeta)}\,.
\eeq
For vanishing spin-spin correlation $\zeta$, this condition
simplifies to
\beq
\al^2 \geq \,\frac{1}{20}\,.
\eeq

The next remark refers to the spin-spin contribution to energy
density (\ref{ep})
\beq
\ep_{\zeta} =\,-\,\frac{\pi}{48}\,\frac{\ga^2}{\ga^2 +
1}\;G\,\rho^2\,(28\,\al^2 - 11)\,\zeta\,.
\eeq
It could result in the ferromagnetic ordering of spins if $\,\al^2
> 11/28\,$. Whether or not this ordering takes place, depends on
the exact relation between $G\rho$ and temperature, both of
which are on the order of magnitude of $m_{\rm{Pl}}$.

\vspace{3mm}

5. One more comment related to equations (\ref{ep}), (\ref{p}).
As mentioned already, according to them, the equation of state, corresponding to the discussed
gravitational four-fermion interaction, is
\beq
p=-\ep.
\eeq
It is rather well-known that this equation of state results in the exponential
expansion of the Universe. Let us consider in this connection our problem.

We assume that the Universe is homogeneous and isotropic, and thus
is described by the well-known Friedmann-Robertson-Walker
(FRW) metric
\beq\label{fr}
ds^2 =dt^2 - a(t)^2 [dr^2 + f(r)(d\theta^2 +\sin^2\theta\,
d\phi^2)];
\eeq
here $f(r)$ depends on the topology of the Universe as a whole:
\[
f(r) = r^2, \hspace{3mm} \sin^2 r, \hspace{3mm} \sinh^2 r
\]
for the spatial flat, closed, and open Universe, respectively.
As to the function $a(t)$, it depends on the equation of state of
the matter.

The Einstein equations for the FRW metric (\ref{fr}) reduce to
\beq\label{H}
H^2 \equiv \left(\frac{\dot{a}}{a}\right)^2 = \frac{8\pi G \ep}{3} - \frac{k}{a^2}\,,
\eeq
\beq\label{a}
\frac{\ddot{a}}{a} = - \frac{4\pi G}{3} \,(\ep + 3 p\,).
\eeq
They are supplemented by the covariant continuity equation, which can be written as follows:
\beq\label{e}
\dot{\ep} + 3\,H (\ep + p\,)=0; \hspace{5mm} H = \frac{\dot{a}}{a}\,.
\eeq
For the energy-momentum tensor (\ref{ep}), (\ref{p}), dominating
on the Planck scale, and resulting in $\ep = - p$, this last equation
reduces to
\beq\label{e1}
\dot{\ep}=0, \hspace{5mm} \rm{or} \hspace{5mm} \ep = \rm{const}.
\eeq
In its turn, equation (\ref{a}) simplifies to
\beq\label{a1}
\frac{\ddot{a}}{a} = \frac{8\pi G \ep}{3} = \rm{const}.
\eeq
In this way, we arrive at the following expansion law:
\beq\label{a2}
a \sim \exp(Ht),\hspace{5mm} \mathrm{where} \hspace{5mm} H = \sqrt{\frac{8\pi G \ep}{3}} = \rm{const}
\eeq
(as usual, the second, exponentially small, solution of eq. (\ref{a1}) is neglected here).

Thus, the discussed gravitational four-fermion interaction results in the inflation
on the Planck scale.

\section*{Acknowledgments}

\hspace{5mm} I am grateful to D.I. Diakonov, A.D. Dolgov, A.A.
Pomeransky, and A.S.~Rudenko for useful discussions.

The investigation was supported in part by the Russian Ministry of
Science, by the Foundation for Basic Research through Grant No.
11-02-00792-a, by the Federal Program "Personnel of Innovational
Russia" through Grant No. 14.740.11.0082, and by the Grant of the
Government of Russian Federation, No. 11.G34.31.0047.

\section*{References}

1. T.W.B. Kibble, {\it J. Math. Phys.} {\bf 2} (1961) 212.\\
2. S. Holst, {\it Phys. Rev.} {\bf D53} (1996) 5966;
gr-qc/9511026.\\
3. I.B. Khriplovich and R.V. Korkin, {\it J. Exp. Theor. Phys.}
{\bf 95} (2002) 1;\\ \indent gr-qc/0112074.\\
4. G.D. Kerlick, {\it Phys. Rev.} {\bf D12}, 3004 (1975).\\
5. A. Perez and C. Rovelli, {\it Phys. Rev.} {\bf D73}, 044013
(2006); gr-qc/0505081.\\
6. L. Freidel, D. Minic, and T. Takeuchi, {\it Phys. Rev.} {\bf
D72}, 104002 (2005);\\ \indent hep-th/0507253.\\
7. G. de Berredo-Peixoto, L. Freidel, I.L. Shapiro, and C.A. de
Souza;\\ \indent gr-qc/12015423.\\
8. I.B. Khriplovich, {\it Phys. Lett.} {\bf B709}, 111 (2012);
gr-qc/12014226.\\
9. L.D. Landau and E.M. Lifshitz, {\it The Classical Theory of
Fields}\\ \indent (Butterworth - Heinemann, 1975).

\end{document}